\documentclass[pra,twocolumn,aps,letterpaper, preprintnumbers,superscriptaddress]{revtex4}

\usepackage{hyperref}
\usepackage{verbatim}
\usepackage{amsmath}
\usepackage{latexsym}
\usepackage{revsymb}
\usepackage{yfonts}
\usepackage{ifthen}
\usepackage{tcolorbox}
\usepackage{natbib}
\usepackage{amsfonts}
\usepackage{amsmath}
\usepackage{amssymb}
\usepackage{amsthm}
\usepackage{graphicx}
\usepackage{bm}
\usepackage{bbm}
\usepackage{epsfig,color,amssymb}
\usepackage{subfigure}
\usepackage{amsfonts}
\usepackage{amscd}
\usepackage{amsmath}
\usepackage{multirow}
\usepackage{chemarrow}
\usepackage{dcolumn}
\usepackage{bm}
\usepackage{graphicx}
\usepackage{enumerate}
\usepackage{epsfig}
\usepackage{subfigure}
\usepackage{xcolor}
\usepackage{multirow}
\usepackage{ulem}
\usepackage{braket}
\usepackage{comment}
\usepackage{enumitem}
\usepackage{amsthm}

\renewcommand{\emph}[1]{\textit{#1}}

\begin{document}
	
	\title{Secure and practical multiparty quantum digital signatures}
	
	\author{Chen-Xun Weng}
	\affiliation{National Laboratory of Solid State Microstructures, School of Physics and Collaborative Innovation Center of Advanced Microstructures, Nanjing University, Nanjing 210093, China}
	
	\author{Yu-Shuo Lu}
	\affiliation{National Laboratory of Solid State Microstructures, School of Physics and Collaborative Innovation Center of Advanced Microstructures, Nanjing University, Nanjing 210093, China}
	
	\author{Rui-Qi Gao}
	\affiliation{National Laboratory of Solid State Microstructures, School of Physics and Collaborative Innovation Center of Advanced Microstructures, Nanjing University, Nanjing 210093, China}
	
	\author{Yuan-Mei Xie}
	\affiliation{National Laboratory of Solid State Microstructures, School of Physics and Collaborative Innovation Center of Advanced Microstructures, Nanjing University, Nanjing 210093, China}
	
	\author{Jie Gu}
	\affiliation{National Laboratory of Solid State Microstructures, School of Physics and Collaborative Innovation Center of Advanced Microstructures, Nanjing University, Nanjing 210093, China}
	
	\author{Chen-Long Li}
	\affiliation{National Laboratory of Solid State Microstructures, School of Physics and Collaborative Innovation Center of Advanced Microstructures, Nanjing University, Nanjing 210093, China}
	
	\author{Bing-Hong Li}
	\affiliation{National Laboratory of Solid State Microstructures, School of Physics and Collaborative Innovation Center of Advanced Microstructures, Nanjing University, Nanjing 210093, China}
	
	\author{Hua-Lei Yin}
	\email{hlyin@nju.edu.cn}
	\affiliation{National Laboratory of Solid State Microstructures, School of Physics and Collaborative Innovation Center of Advanced Microstructures, Nanjing University, Nanjing 210093, China}
	
	\author{Zeng-Bing Chen}
	\email{zbchen@nju.edu.cn}
	\affiliation{National Laboratory of Solid State Microstructures, School of Physics and Collaborative Innovation Center of Advanced Microstructures, Nanjing University, Nanjing 210093, China}
	
	\date{\today}

\begin{abstract}
Quantum digital signatures (QDSs) promise information-theoretic security against repudiation and forgery of messages. Compared with currently existing three-party QDS protocols, multiparty protocols have unique advantages in the practical case of more than two receivers when sending a mass message. However, complex security analysis, numerous quantum channels and low data utilization efficiency make it intractable to expand three-party to multiparty scenario. Here, based on six-state non-orthogonal encoding protocol, we propose an effective multiparty QDS framework to overcome these difficulties. The number of quantum channels in our protocol only linearly depends on the number of users. The post-matching method is introduced to enhance data utilization efficiency and make it linearly scale with the probability of detection events even for five-party scenario. Our work compensates for the absence of practical multiparty protocols, which paves the way for future QDS networks.
\end{abstract}

\maketitle

\section{Introduction}

Digital signature can verify the authenticity of digital messages and has been widely applied in e-mail, e-commerce and software distribution~\cite{diffie1976new}. As e-commerce becomes more and more significant in modern society, the need of unconditionally secure digital signatures against hacking attacks has arisen. Classical digital signatures offer security based on the computational complexity of mathematical problems~\cite{rivest1978method,10000100771,1057074,johnson2001elliptic}. However, the task of rapidly solving these mathematical problems becomes feasible when a quantum computer is available~\cite{shor1994algorithms,doi:10.1137/S0036144598347011,nielsen2001quantum,arute2019quantum,Zhong1460}. Fortunately, quantum digital signatures (QDSs) can offer information-theoretic security relying on quantum mechanics against adversaries who are supposed to have unbounded ability allowed by physics.   

The first QDS protocol was proposed in 2001~\cite{gottesman2001quantum}, but there are some challenging requirements, such as secure quantum channels and long-term quantum memory. After that, the requirement of quantum memory was removed by converting the quantum signatures to classic information through quantum measurements, which makes QDS closer to real implementation~\cite{clarke2012experimental,dunjko2014quantum, collins2014realization,wallden2015quantum,Croal:2016:Free}. Whereas, the security analyses of early protocols still rely on secure quantum channels where there is no eavesdropping. To further improve practicality, two independent QDS protocols without secure quantum channels were proposed and proved to be secure, which are based on non-orthogonal encoding~\cite{yin2016practical} and orthogonal encoding~\cite{PhysRevA.93.032325}, respectively. After these two protocols, numerous excellent achievements of QDS have been made theoretically and experimentally~\cite{puthoor2016measurement,PhysRevA.94.042314,PhysRevA.95.032334,collins2017experimental,yang2017theoretically,PhysRevA.95.042338,roberts2017experimental,zhang2018proof,Thornton:2019:CV,An:19,Qu:19,zhang2019high,ding2020280,zhang2020practical,zhang2020twin,PhysRevA.103.012410,PhysRevX.11.011038}. Protocols based on orthogonal encoding~\cite{PhysRevA.93.032325,puthoor2016measurement,zhang2020twin} need additional symmetrization step which results in extra channels. Recently, drawing on the experience of the four-state Scarani-Acin-Ribordy-Gisin 2004 quantum key distribution (SARG04 QKD) protocol~\cite{PhysRevLett.92.057901,PhysRevA.73.012337,PhysRevA.73.010302,Jeong_2014,RevModPhys.81.1301,yin2016security}, a post-matching QDS protocol has been proposed based on non-orthogonal encoding~\cite{Lu:21}. It does not require additional symmetrization step and also achieves better performance than the original protocol~\cite{yin2016practical}.

\begin{figure*}[t]
	\centering
	\includegraphics[width=0.92\linewidth]{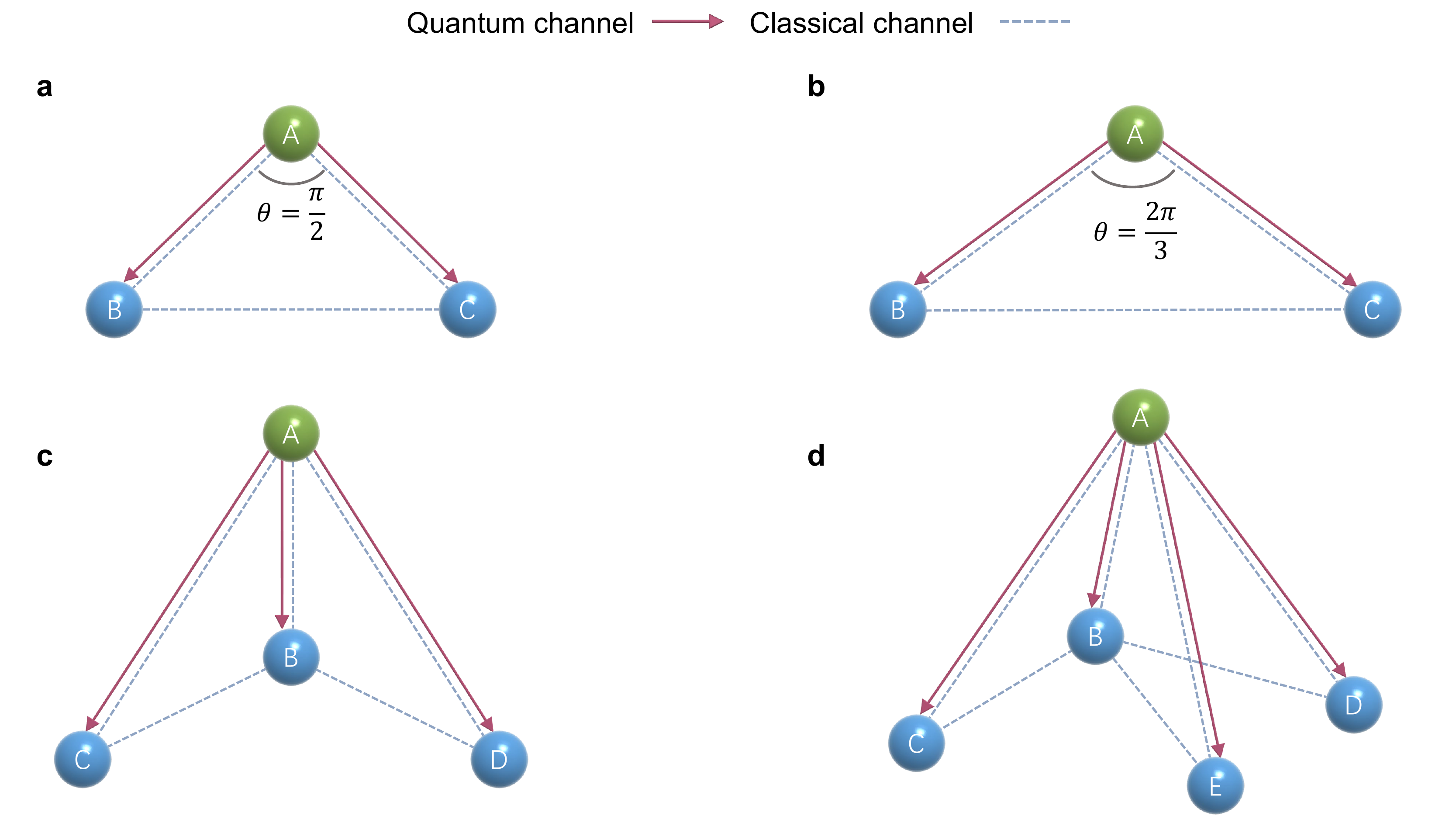}
	\caption{Schematic diagrams of three-party, four-party and five-party protocol. The red line represents insecure quantum channel and the blue dash line is authenticated classical channel. $\theta$ is the angle between Alice-Bob and Alice-Charlie. \textbf{a.} Three-party protocol with $\theta=\frac{\pi}{2}$. \textbf{b.} Three-party protocol with $\theta=\frac{2\pi}{3}$. \textbf{c.} Four-party protocol. \textbf{d.} Five-party protocol.}\label{fig.1}
\end{figure*}

Current QDS protocols mostly focus on three-party communication since the protocol involving more than two recipients will raise three major concerns. The first one is the increased number of quantum communication channels~\cite{longmate2020signing}. When extending orthogonal encoding protocol to multiparty scenarios, each pair of participants requires a quantum communication channel to symmetrize their secret keys. The $M$-party orthogonal encoding protocol requires $M(M-1)/2$ quantum channels. As $M$ increases, it becomes more complex and less practical to implement. The second one is the poor data utilization efficiency leading to low signature rate if we expand the original non-orthogonal encoding protocol to multiparty scenarios. This is because the original non-orthogonal encoding protocol only consider coincidence detection events as valid events. For $M$-party protocol, it requires all detectors of $M-1$ recipients click. Let $\eta$ be the probability that one recipient detector clicks. When the signer sends $N$ quantum states to recipients, there are only $N\eta^{M-1}$ valid events, which is far from enough to perform multiparty protocols. Besides, complex security analysis is also a difficulty to be overcome because there exists a situation where some participants collude with each other to deceive others~\cite{cai2019cryptanalysis}. Although the security analysis of multi-party quantum digital signature schemes based on orthogonal encoding has made progress~\cite{DBLP:journals/qic/ArrazolaWA16,_AH_N_2016}, it does not give an exact example and concrete simulation results.

In this paper, we propose a six-state three-party QDS protocol to enhance performance of signature rate and stability with the help of its higher bit error rate threshold compared with~\cite{Lu:21}. Furthermore, considering that three-photon or even four-photon components of six-state protocol can be used for the secure key, we extend this six-state protocol to four-party and five-party scenarios and overcome difficulties above,  as shown in Fig.~\ref{fig.1}. According to our multiparty QDS framework, we simulate the performance of our three-party, four-party and five-party QDS and give a comparison among them. It is the first practical multiparty QDS framework and we provide security analysis.

\begin{table}[h!]
	\begin{center}
		\caption{Brief description of M-party}
		\begin{tabular}{l|p{0.32\textwidth}}% <-- Alignments: 1st column left, 2nd middle and 3rd right, with vertical lines in between
			\hline
			\hline
			\textbf{Key generation}  &Alice prepares $M-1$ different quantum state sequences and sends them to $M-1$ recipients respectively. All recipients measure quantum states they received in the X, Y or Z basis at random and announce all click events. All participants discard no-click data and keep click data to form their own strings. After that, they perform post-matching process and encode their processed data strings by our rule.\\
			\hline
			\textbf{Estimation}  &Alice informs any verifier to randomly select a certain proportion of strings as test bits. The verifier announces the location of test bits and asks Alice to publicly announce the data information of test bits. The recipients estimate the mismatching rate of conclusive results between their own string and Alice's string. \\
			\hline
			\textbf{Messaging}   & To sign one-bit message, Alice sends her own untested data string to Bob. Whether Bob accepts it depends on Bob's mismatching rate of conclusive results. If Bob accepts, Bob forwards it to all verifiers respectively. Whether verifiers accept it depends on their own mismatching rate. All participants negotiate whether aborting the protocol according to the majority voting principle.	 \\
			\hline
			\hline
		\end{tabular}
	\end{center}
\end{table}

\section{Protocol description}

Let us start by the common notation. The `signer' Alice assigns any one of recipients as `authenticator' and other recipients become `verifiers' automatically. For simplicity, we always let Bob become the authenticator and other recipients become verifiers automatically. Note that in this article we consider the symmetric situation where the fiber lengths between Alice and any one of recipients we mentioned in the following are equal. We will introduce detailed security analysis in Methods. In Table 1, we give a concise description of the framework.

\noindent
\textbf{Three-party protocol.}
We take three-participant scenario as an example and describe all processes in detail. Alice chooses Bob as authenticator and Charlie becomes verifier. 
In our protocol, there are insecure quantum channels connecting Alice with Bob and Alice with Charlie. Moreover, there are authenticated classical channels between any two of three participants.
There are six quantum states: $\left|+x\right\rangle$, $\left|-x\right\rangle$,
$\left|+y\right\rangle$, $\left|-y\right\rangle$, $\left|+z\right\rangle$, $\left|-z\right\rangle$. $\left|\pm x\right\rangle$ are the eigenstates of Pauli X operator. $\left|\pm y\right\rangle$ are the eigenstates of Pauli Y operator. $\left|\pm z\right\rangle$ are the eigenstates of Pauli Z operator.
These six states can be arranged into the 12 sets: $\{\ket{\omega_1x},\ket{\omega_2y}\}$, $\{\ket{\omega_3y},\ket{\omega_4z}\}$ and $\{\ket{\omega_5z},\ket{\omega_6x}\}$, where $\omega_1$, $\omega_2$, $\omega_3$, $\omega_4$, $\omega_5$, $\omega_6$ $\in$ $\{+,-\}$. The first state in each set is encoded with bit value 0 and the second is encoded with bit value 1.

\begin{figure}[t]
	\centering
	\includegraphics[width=9cm]{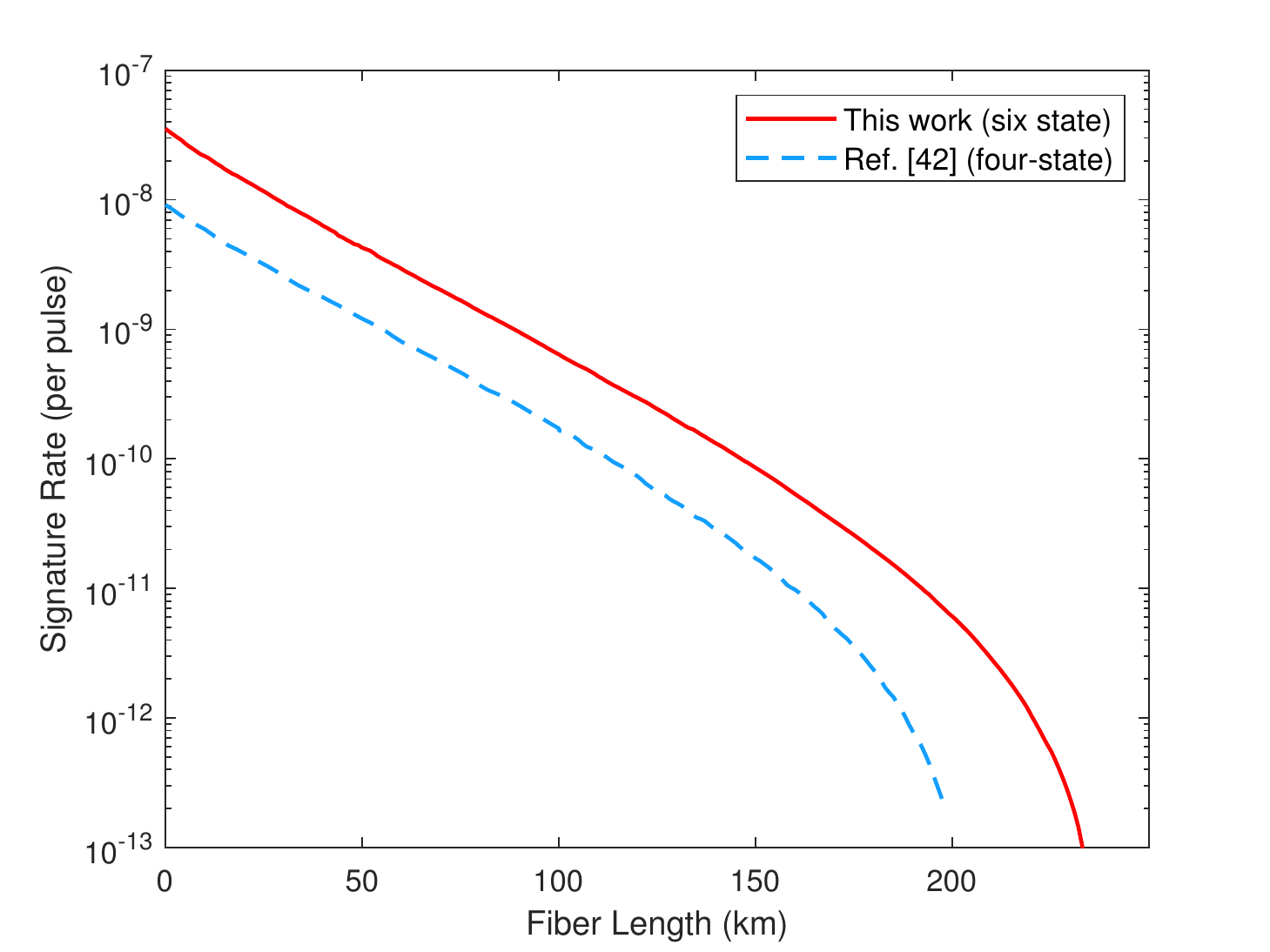}
	\caption{Comparison of performance between our three-party protocol and four-state protocol~\cite{Lu:21}. The detection efficiency is $93\%$. The dark counting rate is $1\times10^{-7}$. The basis misalignment rate is $0.50\%$. The loss coefficient of fiber is 0.16 dB/km. As the fiber length increases, the superiority of our protocol becomes more apparent. The signature rate of our protocol is at least $400\%$ higher than that of~\cite{Lu:21} in this case.}
	\label{fig:2}
\end{figure}

There are three steps in our QDS protocol: key generation, estimation and messaging. 

In the key generation step, Alice uses phase-randomized weak coherent-state source to prepare the six states. For each possible message $m=0$ and $m=1$, Alice prepares two different unrelated sequences of quantum states $A_{B,m}$ and $A_{C,m}$ with length $N$ respectively. Each state is randomly selected from the six states with the same probability by Alice. We denote the light intensity as $\lambda$ ($\lambda\in\left\{\mu, \nu, 0\right\}$). Each quantum state is prepared with the intensity $\mu$, $\nu$ or $0$ and the corresponding possibility $p_{\mu}$, $p_{\nu}$, $p_{0}$ respectively. Alice sends sequences $A_{B,0}$, $A_{B,1}$ to Bob and $A_{C,0}$, $A_{C,1}$ to Charlie through insecure quantum channels. Bob and Charlie receive the sequences and then measure each quantum state in the X, Y or Z basis at random. Bob (Charlie) announces all the click events in $A_{B,m}$ ($A_{C,m}$) through authenticated classical channel, denoted as $S_{B,m}$ ($S_{C,m}$). Afterwards, Alice discards no-click data and keeps click data of length $n$ to form strings $S_{AB,m}$ and $S_{AC,m}$. Alice publicly announces intensity information of all pulses and all three participants divide their remaining data strings into $\mu$-string, $\nu$-string and $0$-string according to the intensity information. For instance, Bob divides $S_{B,m}$ into $S_{B,m}^{\mu}$, $S_{B,m}^{\nu}$ and $S_{B,m}^{0}$ according to the public intensity information.

The three participants perform post-matching method~\cite{Lu:21}. Alice takes the order of quantum states in $S_{AB,m}^{\lambda}$ as a reference and changes the order of $S_{AC,m}^{\lambda}$ to make it same as the order of $S_{AB,m}^{\lambda}$. Alice requests Charlie to change the order of $S_{C,m}^{\lambda}$ into the same order.
For instance, if $S_{AB, m}$$=$ $\{s_{AB,m}^1$, $s_{AB,m}^2$, $s_{AB,m}^3$, $s_{AB,m}^4$, $s_{AB,m}^5$, $s_{AB,m}^6\}$$=$$\{\ket{+x}$, $\ket{-x}$, $\ket{+y}$, $\ket{-y}$, $\ket{+z}$, $\ket{-z}\}$, $S_{AC, m}$$=$$\{s_{AC,m}^1$, $s_{AC,m}^2$, $s_{AC,m}^3$, $s_{AC,m}^4$, $s_{AC,m}^5$, $s_{AC,m}^6\}$$=$$\{\ket{+y}$, $\ket{+z}$, $\ket{-x}$, $\ket{-y}$, $\ket{-z}$, $\ket{+x}\}$, Alice changes initial $S_{AC,m}$ into $S'_{AC,m}$, where $S'_{AC,m}$$=$$\{s_{AC, m}^{6}$, $s_{AC, m}^{3}$, $s_{AC, m}^{1}$, $s_{AC, m}^{4}$, $s_{AC, m}^{2}$, $s_{AC, m}^{5}\}$. She also informs Charlie to change the order of elements in $S_{C, m}$ into $S'_{C, m}$, where $S'_{C, m}$$=$$\{s_{C, m}^{6}$, $s_{C, m}^{3}$, $s_{C, m}^{1}$, $s_{C, m}^{4}$, $s_{C, m}^{2}$, $s_{C, m}^{5}\}$. Note that $S_{C,m}$ is the measurement result of $S_{AC,m}$, so $S'_{C,m}$ is the measurement result of $S'_{AC,m}$. As a result, although Alice sends two different quantum-state sequences, after post-matching process, two identical sequences $S_{AB,m}$ and $S'_{AC,m}$ are obtained by Bob and Charlie respectively.

We illustrate our rule to generate logic bits as follows. For each quantum state sent, Alice randomly chooses one of 12 sets so that the state she sent is one of the two states in the set. Then she assigns the quantum state to this set. When the measurement outcome is orthogonal to any quantum state of the assigned set, the receiver gets a conclusive result encoded with logic bit 0 (the first state) or logic bit 1 (the second state). Otherwise, the receiver gets an inconclusive result denoted as '$\bot $'. They do not announce whether the results are conclusive or inconclusive. Following the rule, all of three participants encode their data strings with  $K_{A, m}^\lambda$, $K_{B, m}^\lambda$ and $K_{C, m}^\lambda$ respectively. The function of binary logic is to quantify the mismatching rate of conclusive results between Alice’s binary encoded data string and each recipient’s binary encoded data string in the estimation step, which is used in the later security analysis.

Here is the example of binary encoding process. The recipients randomly choose X, Y or Z basis to measure each quantum state Alice sent. Alice should publicly announce which of the 12 sets she picked for each state she sent. The set Alice picked should include the state she sent. The recipients will get a conclusive result if any one of the two states in the set is the eigenstate of the basis the recipient chose to measure. For example, Alice sends the state $\ket{+x}$. She will assign it to any one of $\{\ket{+x},\ket{+y}\}$, $\{\ket{+x},\ket{-y}\}$, $\{\ket{+z},\ket{+x}\}$ and $\{\ket{-z},\ket{+x}\}$. When she assigns it to $\{\ket{+x},\ket{+y}\}$ and Bob’s measurement outcome is  $\ket{-y}$ ($\ket{-x}$), Bob will get a conclusive result with logic bit value 0 (1).

In the estimation step, we use superscript $c$ to denote conclusive results, $u$ to denote untested bits, $t$ to denote test bits. The three participants estimate the bit error rate of single-photon pair components with decoy-state method in their $\mu$ strings. Alice announces the information of intensity $\lambda=\nu$ and $\lambda=0$ publicly. Alice informs Charlie to randomly select a certain proportion, denoted as $t$, of $\mu$ strings as test bits. Charlie announces the location of test bits and asks Alice to publicly announce the data information of test bits. Denote the mismatching rate of conclusive results between $K_{A,m}^{t}$ and $K_{B,m}^{t}$ (between $K_{A,m}^{t}$ and $K_{C,m}^{t}$) as $E_{B}^{ct}$ ($E_{C}^{ct}$). Moreover, Bob and Charlie calculate the proportion of conclusive results in $K_{B,m}$ and $K_{C,m}$ respectively, denoted as $P_{B}^{c}$ and $P_{C}^{c}$. If either of them deviates greatly from the ideal value $\frac{1}{6}$, they also abort the protocol. Afterwards, all of them throw away the test bits and conserve the untested bits of $\mu$ strings with remaining length $(1-t)n_{\mu}$. 

\begin{figure}[t]
	\centering
	\includegraphics[width=9cm]{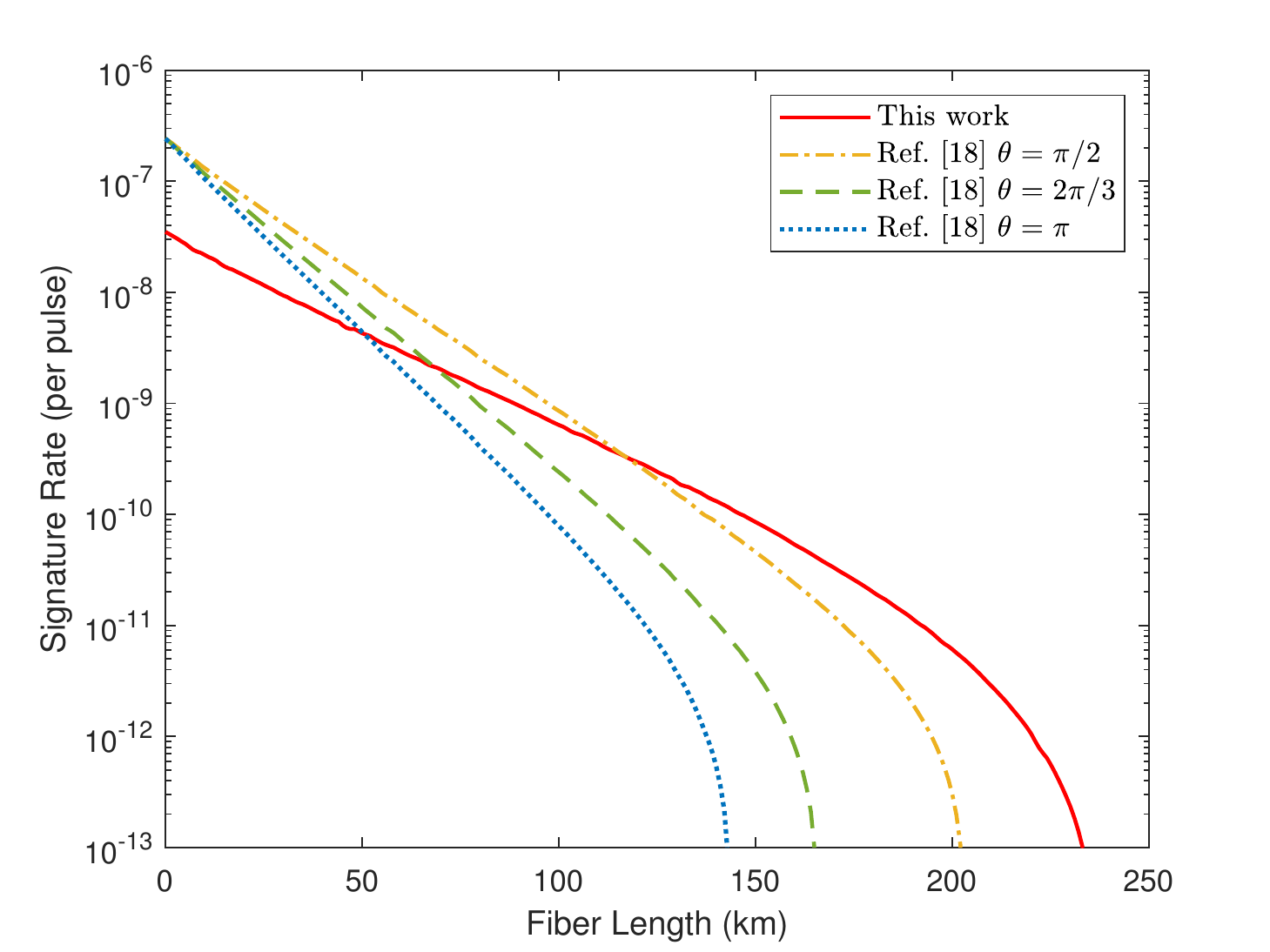}
	\caption{Comparison of performance between our three-party protocol and orthogonal encoding protocol~\cite{PhysRevA.93.032325}. We simulate two protocols under the same experimental parameters. The signature rate of our protocol is lower at short distance. However, it decays more slowly than orthogonal encoding protocol and shows better performance especially at long distance. In this case, our protocol has a longer transmission distance.}
	\label{fig:3}
\end{figure}
In the messaging step, to sign one-bit message $m$, Alice sends $\left\{m,K_{A,m}^{u}\right\}$ to Bob. Bob checks the mismatching rate of conclusive results $E_{B}^{cu}$ between $K_{A,m}^{u}$ and $K_{B,m}^{u}$. If $E_{B}^{cu}\le T_{a}$ ($T_{a}$ is the authentication security threshold), Bob accepts the message. Otherwise, he rejects the message and aborts the protocol. When Bob accepts the message from Alice, he forwards $\left\{ m,K_{A,m}^{u} \right\}$ to verifier Charlie. After that, Charlie checks the mismatching rate of conclusive results $E_{C}^{cu}$ between $K_{A,m}^{u}$ and $K_{C,m}^{u}$. If $E_{C}^{cu}\le T_{v}$ ($T_{v}$ is the verification security threshold), Charlie accepts the message. Otherwise, Charlie rejects the message and aborts the protocol.

\begin{figure}[t]
	\centering
	\includegraphics[width=9cm]{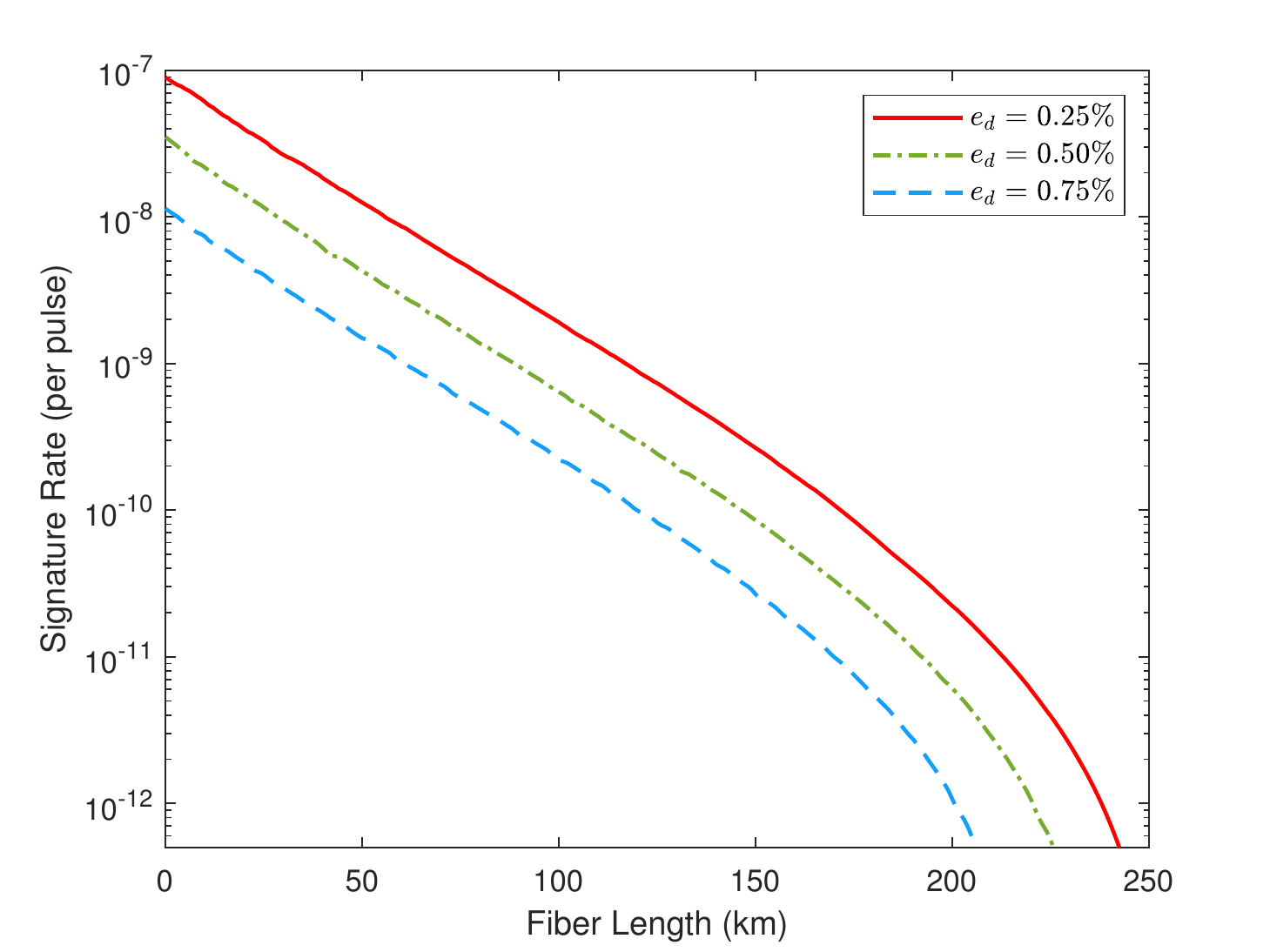}
	\caption{Optimal signature rate of three-party protocol with the same dark counting rate $p_d=1\times10^{-7}$ under different basis misalignment rate.}
	\label{fig:4}
\end{figure}

We define the signature rate as $R:=\frac{1}{2N}$, where $2N$ is the minimum number of pulses required to securely sign a one-bit message. We define that it is secure enough to sign a 1-bit message when the robustness $\varepsilon _{rob}$, the probability of successful forgery $\varepsilon_{for}$, the probability of successful repudiation $\varepsilon_{rep}$, the failure probability of the Chernoff bound $\epsilon_1$ and the failure probability of random sampling without replacement $\epsilon_2$ do not exceed their thresholds respectively.

\begin{figure}[t]
	\centering
	\includegraphics[width=9cm]{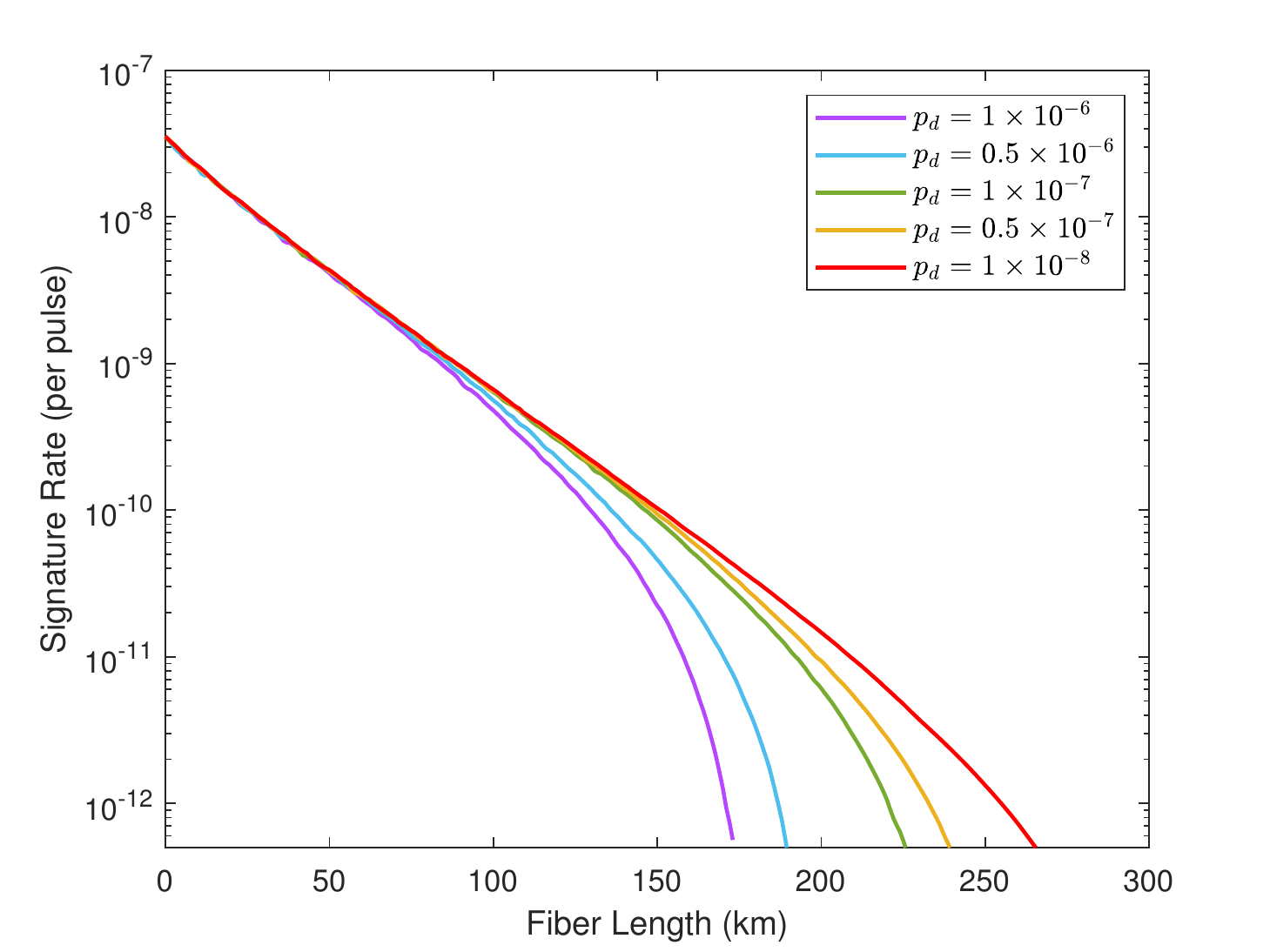}
	\caption{Optimal signature rate of three-party protocol with the same basis misalignment rate $e_d=0.5\%$ under different dark counting rates.}
	\label{fig:5}
\end{figure}

As shown in Fig.~\ref{fig:2}, we simulate the four-state protocol of~\cite{Lu:21} to compare it with our six-state protocol. The performance of our protocol is better than that of~\cite{Lu:21}. For example, when the fiber length is 150 km, the original four-state protocol requires about $2.93\times10^{10}$ pulses to sign a one-bit message. However, under the same conditions our six-state protocol only needs about $5.85\times10^{9}$ pulses. When the fiber length is 150 km, the signature rate of our protocol is approximately $500\%$ higher than four-state . Moreover, we also simulate the performance of orthogonal encoding based protocol with symmetrization step in~\cite{PhysRevA.93.032325} to compare with our protocol as shown in Fig.~\ref{fig:3}. We denote the angle between Alice-Bob and Alice-Charlie as $\theta$. Denote the distance between Alice and Bob (Charlie) as $D_{AB}$ ($D_{AC}$) and the distance between Bob and Charlie as $D_{BC}$. For a symmetric case, $D_{AB}$$=$$D_{AC}$ and $D_{BC}=2D_{AB}\sin \theta/2$. $D_{BC}$ increases as $\theta$ gets larger. When $\theta$ is close to $\pi$ , the transmission distance of QKD ($D_{BC}$) increases much faster than $D_{AB}$. Detailed information can be found in Ref.~\cite{Lu:21}. Define the effective signature rate as $R_{eff}:=\max\{\frac{1}{2N},\frac{R_{QKD}}{6L}\}$, where $L$ is the length of generated key and $R_{QKD}$ is the secret key rate of QKD. $R_{QKD}$ is simulated by the key rate formula of~\cite{PhysRevA.89.022307}. We simulate three cases of $\theta=\pi$, $\theta=\frac{2\pi}{3}$ and $\theta=\frac{\pi}{2}$. Our protocol has a longer transmission distance and greater performance of signature rate especially at long distance in these cases where the signature rate of our protocol decays more slowly. We also simulate our protocol's performance under different dark counting rates and different basis misalignment rates as shown in Fig.~\ref{fig:4} and Fig.~\ref{fig:5} respectively. From two figures, we can see that our protocol shows obviously high error rate tolerance and stability against noise.

\bigskip
\noindent
\textbf{Four-party QDS protocol.}
In our four-party protocol, there are `signer' Alice, `authenticator' Bob, `verifier' Charlie and `verifier' David. Their positions are shown in Fig.~\ref{fig.1}c. The operation among Alice, Bob and Charlie are the same as three-party protocol, which we will not describe in detail here. We focus on the difference due to the new participant `verifier' David instead.

For each possible message $m=0$ and $m=1$, following the rule of generating logic bits, David encodes his data strings with $K_{D,m}^{\lambda}$ in the key generation step.

In the estimation step, the four participants estimate the bit error rate of triple-photon components with decoy-state method in their $\mu$ strings. Alice announces all information of intensity $\lambda=\nu$ and $\lambda=0$. Then Alice informs any one of verifiers to randomly select a proportion of $\mu$ strings as test bits. All verifiers announce the location of their test bits respectively and request Alice to announce the data information of test bits publicly. Denote the mismatching rate of conclusive results between $K_{A,m}^{t}$ and $K_{B,m}^{t}$ as $E_{B}^{ct}$, the mismatching rate of conclusive results between $K_{A,m}^{t}$ and $K_{C,m}^{t}$ as $E_{C}^{ct}$ and the mismatching rate of conclusive results between $K_{A,m}^{t}$ and $K_{D,m}^{t}$ as $E_{D}^{ct}$. Moreover, Bob, Charlie and David calculate the proportion of conclusive results in $K_{B,m}$, $K_{C,m}$ and $K_{D,m}$ respectively, denoted as $P_{B}^{c}$, $P_{C}^{c}$ and $P_{D}^{c}$.

\begin{figure}[t]
	\centering
	\includegraphics[width=9cm]{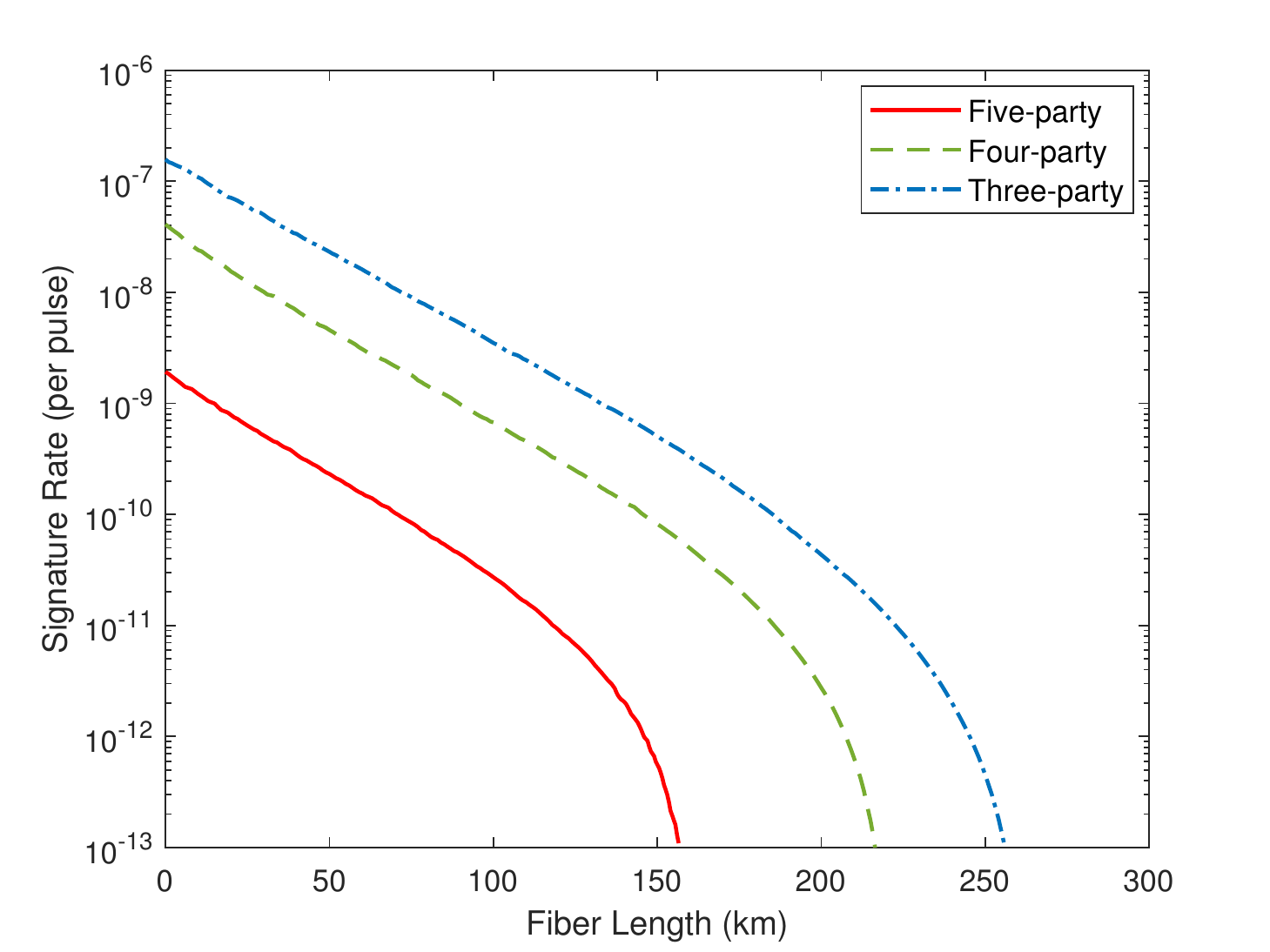}
	\caption{The performance of three-party, four-party and five-party protocol under the same basis misalignment rate $e_d=0.1\%$. With the help of post-matching method, extending three-party to four-party or even five-party scenario is feasible because our data utilization efficiency is highly improved.}
	\label{fig:6}
\end{figure}

To sign one-bit message $m$, Alice sends $\left\{m,K_{A,m}^{u}\right\}$ to Bob. Bob checks the mismatching rate of conclusive results $E_{B}^{cu}$ between $K_{A,m}^{u}$ and $K_{B,m}^{u}$. If $E_{B}^{cu}\le T_{a}$, Bob accepts the message. Otherwise, he rejects the message and aborts the protocol. When Bob accepts the message from Alice, he forwards $\left\{ m,K_{A,m}^{u} \right\}$ to Charlie and David respectively. After that, Charlie checks the mismatching rate of conclusive results $E_{C}^{cu}$ between $K_{A,m}^{u}$ and $K_{C,m}^{u}$. If $E_{C}^{cu}\le T_{Cv}$ ($T_{Cv}$ is the verification security threshold of Charlie), Charlie accepts the message. Otherwise, Charlie rejects the message. David checks the mismatching rate of conclusive results $E_{D}^{cu}$ between $K_{A,m}^{u}$ and $K_{D,m}^{u}$. If $E_{D}^{cu}\le T_{Dv}$ ($T_{Dv}$ is the verification security threshold of David), David accepts the message. Otherwise, David rejects the message. Either of Charlie and David rejects the message means that the protocol will be aborted. All participants negotiate whether aborting the protocol or not according to the majority voting principle.

\bigskip
\noindent
\textbf{Five-party QDS protocol.}
When it comes to our five-party protocol, there are five participants `signer' Alice, `authenticator' Bob, `verifier ' Charlie,  `verifier' David and `verifier' Emery. Their positions are shown in Fig.~\ref{fig.1}d. The processes among Alice, Bob, Charlie and David are the same as the four-party protocol. We only focus on the operation involving Emery here. 

In the key generation step, Emery encodes his data strings with $K_{E,m}^{\lambda}$ following the process as we described above.

In the estimation step, the five participants estimate the bit error rate of four-photon component with decoy-state method in their $\mu$ strings. Alice announces the information of intensity $\lambda=\nu$ and $\lambda=0$. Alice informs any one verifier to randomly select a certain proportion of $\mu$ strings as test bits. The participants estimate their mismatch rate of conclusive results. Denote the mismatching rate of conclusive results between $K_{A,m}^{t}$ and $K_{E,m}^{t}$ as $E_{E}^{ct}$. Moreover, Emery calculates the proportion of conclusive results in $K_{E,m}$, denoted as $P_{E}^{c}$. If any one of Bob, Charlie, David and Emery deviates greatly from the ideal value $\frac{1}{6}$, they abort the protocol. Afterwards, all of them throw away the test bits and keep the untested bits of $\mu$ strings with remaining length $(1-t)n_{\mu}$. 

In the messaging step, Alice sends $\left\{m,K_{A,m}^{u}\right\}$ to Bob in order to sign one-bit message $m$. Bob checks $E_{B}^{cu}$. If $E_{B}^{cu}\le T_{a}$, Bob accepts the message. Otherwise, he rejects the message and aborts the protocol. When Bob accepts the message from Alice, he forwards $\left\{ m,K_{A,m}^{u} \right\}$ to Charlie, David and Emery respectively. Emery checks the mismatching rate of conclusive results $E_{E}^{cu}$ between $K_{A,m}^{u}$ and $K_{E,m}^{u}$. If $E_{E}^{cu}\le T_{Ev}$ ($T_{Ev}$ is the verification security threshold of Emery), Emery accepts the message. Otherwise, Emery rejects. All participants negotiate whether aborting the protocol or not according to the majority voting principle.

\section{Security analysis}

Our security analysis follows~\cite{yin2016practical} and~\cite{Lu:21}. We build the framework for $M$-party ($M$=$3, 4, 5$) protocol about three security criteria: robustness, security against forgery and security against repudiation. 
Here, we apply majority voting principle to solve dispute. For four-party protocol, there are at most one dishonest participant. Any two of participants making the wrong decision leads to successful attack. For five-party protocol, we should consider the colluding attack where there are two dishonest participants. We can assume Emery is a fixed dishonest player and he will collude with the other dishonest participant (Alice or Bob). Emery always unconditionally supports his partner. In other words, Charlie and David must make the same correct decision. This situation is equivalent to the four-party scenario above where there exists only one dishonest participant among Alice, Bob, Charlie and David.

The upper bound and lower bound of expected value of parameter $a$ can be given by a variant of Chernoff bound~\cite{yin2020tight}: $\overline{a}^*=a+\beta+\sqrt{2\beta a +\beta^2}$ and $\underline{a}^*=a-\frac{\beta}{2}-\sqrt{2\beta a +\frac{\beta^2}{4} }$ where $\beta=\ln\frac{1}{\varepsilon_1}$ and $\epsilon_1$ is the failure probability of the Chernoff bound. We use $k$ to denote $k$-photon component, where $k=M-1$.
\paragraph{(1) Robustness}
Robustness ($\varepsilon_{rob}$) represents the probability that the protocol is aborted when the antagonist is inactive. In messaging step, Bob does not accept the message if $E_{B}^{cu}>T_{a}$. We can quantify robustness by random sampling without replacement theorem~\cite{yin2020tight} in finite sample case. 
\paragraph{(2) Security against forgery}
Forgery attack means Bob wishes that more than half of verifiers would accept the forwarded message forged by Bob $\left\{ m, K_{BF} \right\}$. In this case, Bob needs to obtain as much information as he can about quantum states that all verifiers receive, like an eavesdropper in SARG04 QKD protocol~\cite{PhysRevLett.92.057901,PhysRevA.73.010302,yin2016security,PhysRevA.95.032334}.

All positions of recipients are equal. Without loss of generality, we first consider the probability that Charlie is deceived by Bob. We exploit the decoy state method~\cite{PhysRevLett.94.230503,PhysRevLett.94.230504,PhysRevA.72.012326,PhysRevA.72.012322} to estimate the bit error rate $e_{b}$ of $k$-photon component. 

Considering the process where Alice sends pulses to all recipients, we have
\begin{equation}
	s_{C1}^{c\mu^{\ast}}\ge\frac{p_{\mu}e^{-\mu}}{\nu(\mu-\nu)}(\mu^{2}e^{\nu}\frac{\underline{n}_{C\nu}^{c^{\ast}}}{p_{\nu}}-\nu^{2}e^{\mu}\frac{\overline{n}_{C\mu}^{c^{\ast}}}{p_{\mu}}+(\nu^{2}-\mu^{2})\frac{\underline{n}_{C0}^{c^{\ast}}}{p_{0}}),
\end{equation}
where $s_{C1}^{c\mu^{\ast}}$ is the number of conclusive single-photon events in Charlie's $\mu$ string.
\begin{equation}
	s_{Q1}^{\mu^{\ast}}\ge\frac{p_{\mu}e^{-\mu}}{\nu(\mu-\nu)}(\mu^{2}e^{\nu}\frac{\underline{n}_{Q\nu}^{\ast}}{p_{\nu}}-\nu^{2}e^{\mu}\frac{\overline{n}_{Q\mu}^{\ast}}{p_{\mu}}+(\nu^{2}-\mu^{2})\frac{\underline{n}_{Q0}^{\ast}}{p_{0}}),
\end{equation}
\begin{equation}
	s_{Q1}^{\mu^{\ast}}\le\frac{p_{\mu}\mu e^{-\mu}}{\nu}(e^{\nu}\frac{\overline{n}_{Q\nu}^{\ast}}{p_{\nu}}-\frac{\underline{n}_{Q0}^{\ast}}{p_0}),
\end{equation}
where $Q \in  \Omega$ and
\begin{equation}   \Omega=
	\begin{cases}
		\{B\}  &  \text{if M=3,} \\
		\{B,D\} &  \text{if M=4,}\\
		\{B,D,E\} &  \text{if M=5,}
	\end{cases}                
\end{equation}        
and $s_{Q1}^{\mu^{\ast}}$ is the number of single-photon events in Q's $\mu$ strings. Therefore, we have

\begin{equation}\label{eq5}
	s_{Ck}^{c\mu^{\ast}}\ge\underline{s}_{C1}^{c\mu^{\ast}}\times\prod_{Q\in\Omega} \frac{\underline{s}_{Q1}^{\mu^{\ast}}}{\overline{n}_{Q\mu}^{\ast}},
\end{equation}
where $s_{Ck}^{c\mu^{\ast}}$ is the number of events that all recipients receive a single-photon in $\mu$ string and Charlie has a conclusive result simultaneously. For example, when it comes to four-party, $s_{C3}^{c\mu^{\ast}}\ge\underline{s}_{C1}^{c\mu^{\ast}}\times \frac{\underline{s}_{B1}^{\mu^{\ast}}}{\overline{n}_{B\mu}^{\ast}}\times \frac{\underline{s}_{D1}^{\mu^{\ast}}}{\overline{n}_{D\mu}^{\ast}}$.

We also have
\begin{equation}
	t_{C1}^{c\mu^{\ast}}\le\frac{p_{\mu}\mu e^{-\mu}}{\nu}(e^{\nu}\frac{\overline{m}_{C\nu}^{c^{\ast}}}{p_{\nu}}-\frac{\underline{n}_{C0}^{c^{\ast}}}{2p_0}),
\end{equation}
and

where $t_{C1}^{c\mu^{\ast}}$ is the number of single-photon error events of Charlie's conclusive results in $\mu$ string with respect to Alice. We can get
\begin{equation}
	t_{Ck}^{c\mu^{\ast}}\le\overline{t}_{C1}^{c\mu^{\ast}}\times\prod_{Q\in\Omega}\frac{\overline{s}_{Q1}^{\mu^{\ast}}}{\underline{n}_{Q\mu}^{\ast}}
\end{equation}
where $t_{Ck}^{c\mu^{\ast}}$  is the number of events that all recipients receive a single-photon in $\mu$ string, Charlie has a conclusive result and his classic bit mismatches with Alice's.

Therefore, the bit error rate $e_b$ can be given by $e_b=t_{Ck}^{c\mu^{\ast}}/s_{Ck}^{c\mu^{\ast}}$.

The relationship between phase error rate $e_p$ and bit error rate $e_b$~\cite{yin2016practical} in six-state SARG04 protocol is
\begin{equation}\label{eq8}  e_{p}=
	\begin{cases}
		\frac{4-\sqrt{2}}{4}+\frac{3}{2\sqrt{2}}e_{b}  &  \text{if M=3,} \\
		\frac{1}{4}+\frac{3}{4}e_b &  \text{if M=4,}\\
		\min\{xe_b+f(x)\}, \forall x &  \text{if M=5,}
	\end{cases}                
\end{equation} 
where $f(x)=\frac{6-4x+\sqrt{6-12\sqrt{2}x+16x^2}}{12}$.

$E_{BFk}^*$ can be given by $H(E_{BFk}^*)=1-I_{B}=1-H(e_{p}|e_{b})$, where $H(e_{p}|e_{b})$ is the conditional Shannon entropy function, $I_{B}$ is mutual information provided by~\cite{yin2016practical} and $E_{BFk}^*$ is the expected value of minimum mismatching rate of conclusive results of the k-photon component between correct $K_{A,m}^u$ and forged $K_{BF,m}^u$.

We employ Chernoff Bound~\cite{10.1214/aoms/1177729330} and the probability of successful forgery ($\varepsilon_{for}$) can be given by 
\begin{equation}
	\varepsilon_{for}=\exp\left[ -\frac{(E_{BFk}^*-T_{vk})^2}{2E_{BFk}^*} n^{cu}_{k}\right],
\end{equation}
where $T_{vk}=T_vn^{cu}/n^{cu}_{k}$ is the error rate threshold of $k$-photon component, $n^{cu}=(1-t)n_\mu^c$ is the number of conclusive results in $K_{C,m}^u$ and $n^{cu}_{k}=(1-t)s_{Ck}^{c\mu}$ is the number of k-photon component in $K_{C,m}^u$. Note that $\epsilon_{for}$ is determined by the probability of deceiving the most vulnerable recipient. Therefore,
\begin{equation}  T_{v}=
	\begin{cases}
		T_{Cv}  &  \text{if M=3,} \\
		\min\{T_{Cv},T_{Dv}\} &  \text{if M=4,}\\
		\min\{T_{Cv},T_{Dv},T_{Ev}\}&  \text{if M=5.}
	\end{cases}                
\end{equation} 

\paragraph{(3) Security against repudiation}
Alice repudiates successfully when Bob accepts the message and more than half of the verifiers refuse to accept it.
The probability of repudiation $\varepsilon_{rep}$ can be given by
\begin{equation}
	\varepsilon_{rep}=\exp\left[-\frac{\left(A-P_B^cT_a\right)^2}{2A}n^u\right].
\end{equation}
$A$ is the solution of the following equation:
\begin{equation}
	\frac{\left[P_C^cT_v-P_C^c\left(\frac{\overline{\Delta}^{cu}}{n^{cu}}+\frac{A}{P_B^c}\right)\right]^2}{3P_C^c\left(\frac{\overline{\Delta}^{cu}}{n^{cu}}+\frac{A}{P_B^c}\right)}=\frac{(A-P_B^cT_a)^2}{2A}, 
\end{equation} 
with $ P_B^cT_a<A<P_B^c\left(T_v-\frac{\overline{\Delta}^{cu}}{n^{cu}}\right)$. 
$\overline{\Delta}^{cu}$ can be given by
\begin{equation} \overline{\Delta}^{cu}=
	\begin{cases}
		\overline{\Delta}_{BC}^{cu}  &  \text{if M=3,} \\
		\max\{\overline{\Delta}_{BC}^{cu}, \overline{\Delta}_{BD}^{cu}\} &  \text{if M=4,}\\
		\max\{\overline{\Delta}_{BC}^{cu}, \overline{\Delta}_{BD}^{cu},\overline{\Delta}_{BE}^{cu}\} &  \text{if M=5,}
	\end{cases}                
\end{equation} 
where $\overline{\Delta}_{BC}^{cu}$ is the relative Hamming distance between $E_B^{cu}$ and $E_C^{cu}$, $\overline{\Delta}_{BD}^{cu}$ is the relative Hamming distance between $E_B^{cu}$ and $E_D^{cu}$, $\overline{\Delta}_{BE}^{cu}$ is the relative Hamming distance between $E_B^{cu}$ and $E_E^{cu}$.

Therefore, the total security can be given by
\begin{equation} \varepsilon_{tot}=
	\begin{cases}
		11\epsilon_1+\epsilon_2+\varepsilon_{rob}+\varepsilon_{for}+\varepsilon_{rep}  &  \text{if M=3,} \\
		17\epsilon_1+\epsilon_2+\varepsilon_{rob}+\varepsilon_{for}+\varepsilon_{rep} &  \text{if M=4,}\\
		23\epsilon_1+\epsilon_2+\varepsilon_{rob}+\varepsilon_{for}+\varepsilon_{rep} &  \text{if M=5,}
	\end{cases}                
\end{equation} 
where $\epsilon_1$ is the failure probability of the Chernoff bound and $\epsilon_2$ is the failure probability of random sampling without replacement.

In our simulation, we set the security bounds as $\varepsilon_{tot} \le 10^{-9}$, $\varepsilon_{for} \le 10^{-10}$,  $\varepsilon_{rob}\le 10^{-10}$,  $\varepsilon_{rep}\le 10^{-10}$ and $\varepsilon_1=\varepsilon_2$.

Note that as shown in Fig.~\ref{fig:6}, expanding the QDS framework to an increasing number of users implies that the signature rate $R$ decreases more rapidly than just linearly as the number of parties increases. That is because, for $M$-party protocol, only the $M-1$ photon component can be considered to be secure when we consider security against forgery.  That means the addition of a new user requires an extra single photon reducing the efficiency which makes signature rate get lower as we pointed in Eq.~(\ref{eq5}). Additionally, the relationships between phase error rate $e_p$ and bit error rate $e_b$ of $M-1$ photon component in six-state SARG04 protocol are different as shown in Eq.~(\ref{eq8}). That will also influence the signature rate of multiparty QDS.

Furthermore, the increase of system loss and the decrease of detection efficiencies will both lead to the decrease of valid detection events when sending the same number of pulses. That means the statistical fluctuation will increase resulting in the increase of the probability of successful repudiation and forgery. Moreover, the enhancement of security constraint also results in the statistical fluctuation increasing. Therefore, more pulses are required to keep the protocol safe, i.e., the signature rate will be lower.

%%%%%%%%%%%%%%%%%%%%%%%%%%%%% \\\\\  SECURITY PROOF  ///// %%%%%%%%%%%%%%%%%%%%%%%%%%%%%
\section{Conclusion}

In summary, we have presented a practical QDS framework that consists of multiple participants. In our three-party protocol, signature rate, secure transmission distance and error tolerance achieve better performance because of higher error rate threshold. Additionally, as shown in Fig.~\ref{fig:6}, our protocol can be extended to multiparty scenarios with great performance. In our simulation, when the basis misalignment rate $e_d=0.1\%$ and dark counting rate $p_d=1\times 10^{-7}$, our three-party, four-party and five-party QDS protocols can reach the transmission distance of 265 km, 220 km and 156 km respectively. When the fiber length is 150 km, the signature rates of three-party, four-party and five-party are $5.1\times10^{-10}$, $8.3\times10^{-11}$ and $5.6\times10^{-13}$ respectively. As shown in Fig.~\ref{fig:4}, the signature rate does not decrease dramatically as $e_d$ increases, showing the great fault tolerance. For example, when fiber length is 150 km, the signature rates are $2.7\times10^{-10}$, $8.5\times10^{-11}$ and $2.7\times10^{-11}$ under $e_d=0.25\%$, $0.50\%$ and $0.75\%$ respectively.

The insurmountable barrier for original non-orthogonal encoding protocol to realize multiparty QDS protocol is low data utilization efficiency due to the requirement of coincidence detection. But our $M$-party protocol perfectly overcome this barrier because we can highly increase data utilization efficiency from $O(\eta^{M-1})$ to $O(\eta)$ with post-matching method, resulting in pronounced improvement of signature rate. Compared with orthogonal encoding protocol, our multiparty protocols are concise and maneuverable since our $M$-party protocol only needs $M-1$ quantum channels as we shown in Fig.~\ref{fig.1}. The requirement of fewer quantum channels is a noticeable advantage of our QDS framework. 

Also, in our work, we have presented security analysis of generalized multiparty QDS framework. These multiparty QDS protocols promise robustness, security against forgery and security against repudiation. We also solved the complex problem of colluding attack existing in the five-party scenario which never happens in three-party QDS by majority voting. This work provides specific ideas for practical multiparty QDS protocol. It will be interesting to apply ideas of our QDS framework to realize large-scale QDS networks in the near future.

\acknowledgments

We gratefully acknowledge support from the National Natural Science Foundation of China (under Grant No. 61801420); the Key-Area Research and Development Program of Guangdong Province (under Grant No. 2020B0303040001); the Fundamental Research Funds for the Central Universities (under Grant No. 020414380182).

H.-L.Y. and Z.-B.C.conceived the research. C.-X.W., Y.-S.L. and H.-L.Y. designed the protocol and proved its security. All authors contributed to discussing the results and writing the paper. C.-X.W. and Y.-S.L. contributed equally to this work.

\end{document}